# Novel Network Coding-based Techniques for Multi-layer Video Delivery over Multi-hop Wireless testbed

Fatima Amir Hamza[1], Lamia Romdhani[1], Amr Mohamed[1]

[1]Computer Science and Engineering Department, Qatar University

Doha, Qatar

{fatmaa, lamia.romdani, amrm}@qu.edu.qa

*Abstract*— **The actual performance of video delivery over wireless networks is best captured through experimental studies over real testbed, which is crucial to uncover the technical problems and practical challenges of realizing the proposed video delivery techniques. Network coding (NC) provides a paradigm shift to optimize the network resources for enhancing the video quality as a result of delivery over multi-hop wireless networks. In this paper, we propose a novel framework for evaluating the performance of NC-based video delivery over multi-hop wireless networks. Based on an experimental testbed implementation of inter-layer video coding and delivery, we propose a new generalized technique for multi-hop interlayer NC in enhancing video delivery quality. We evaluate the trade-off between complexity of multi-hop NC, and video quality gain. We provide clear recommendations for best design choices of NC-based inter-layer video coding and delivery over multi-hop wireless mesh network considering link conditions and Packet Delivery Ratio (PDR). The results show that using NC improves the layered video performance in single and multi-hop network. The average number of decoded layers when using NC in single hop outperforms that when using NC in multi-hop.**

*Keywords*— Network Coding, layered video, multi-hop, Wireless Mesh Network, testbed[1].

## I. INTRODUCTION

Wireless Mesh Network (WMN) is a collection of mesh routers and mesh nodes [3]. It is dynamically self-organized and self-configured [14]. The video content over wireless devices becomes increasingly popular, and it is one of the essential applications today. Video delivery over WMNs faces many challenges such as supporting QoS for video and granted delivery for video content. Indeed, the quality of the radio links depends on many factors, including, the contention and the interference among nearby nodes, which causes packet loss, the link's available bandwidth, and the traffic load transmitted over the link [16]. NC is used to enhance video performance in every network layer stack e.g. physical layer, network layer and application layer. In our work, we focus on NC features at application layer to enhance multi-layers video delivery by mixing video layers.

NC promotes the combining of several packets by using linear coefficients before sending them to the receiver. Such basic concept elevates NC as a potential new concept that increases the network throughput by communicating more information with fewer packet transmissions that enhances the bandwidth utilization. Moreover, NC improves network security because adversary node can combine packets using secure coefficients to enhance privacy and increase data integrity [5][9]. Furthermore, increasing the number of hops in WMN reduces the video delivery quality and the effective date rate [4]. In this regard, NC can also be leveraged to enhance the video delivery performance and maximize the bandwidth utilization [12].

There are two types of layer-based NC, namely: intra-layer coding, when the coding is performed within the layer, and inter-layer coding, when the coding is performed across layers [13]. Fig. 1 shows a simple example of inter-layer NC scheme. In Fig. 1(a) the Sender (S) is transmitting video packets to the Receiver (R). S uses Multi-Resolution Coding (MRC) [10] to divide video into n layers one basic layer ($L_1$), and $n-1$ enhancement layer($L_2, L_3, ..., L_n$). Using inter-layer network coding, S may combine lower layer $L_1$ with one or more upper layers($L_2, L_3, ..., L_n$). The combination is done by XORing layer packets. If R receives consecutive layers($L_1, L_2, ..., L_i$) successfully, then R can decode all layers less than or equal layer i. if $L_1$ is not received, then R can't decode any upper layers. In Fig. 1(a) the intermediate node (I) only forwards coded layers to node R. If R receives the combined packet i times, then it can decode i layers using XORing. This transmission is an end-to-end (E2E) scheme. In Fig. 1(b), the node I performs decoding and re-encoding, then it transmits the mixed packets to the receiver. This transmission is named hop by hop (HBH) scheme, and it is similar to the E2E scheme per link in the path.

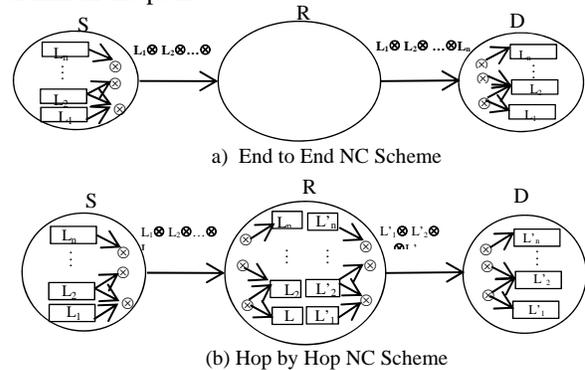

a) End to End NC Scheme

b) Hop by Hop NC Scheme

Fig 1. Inter-Layer Network Coding

To deliver layered video over a single wireless link Koutsonikolas *et al.* [10] proposed an NC mechanism. In this mechanism, the video stream is divided into four layers, one base layer ($L_1$) and three enhancement layers ($L_2, L_3, L_4$). Each layer is divided into multiple packets $P$. The source and the destination respectively encode and decode the different layers using XORing. PERCY is implemented on proxy to send MRC video based on a Canonical Triangular Scheme for Inter-Layer Network Coding. PERCY constructs a Strategy Performance Table (SPT) that stores the expected number of layers decoded under each triangular strategy for a range of PDRs (5% to 100% following a step of 5%). The aim of PERCY is to search and select the optimal vector $(X_1, X_2, X_3, X_4)$ to maximize the expected number of decoded layers at the destination node according to the PDR of the wireless link. PERCY generates coded packets based on optimal vector $(X_1, X_2, X_3, X_4)$, by using the canonical triangular scheme to transmit layered video. The canonical triangular scheme is described in Equation 1. The source sends the combination of $L_1 \otimes L_2 \ldots \otimes L_i, \forall i \in [1,4]$ for $X_i$ times. Hence, when the receiver receives $L_i$ at least $i$ times, it can decode $L_i$ and all lower layers. It should be noted that the performance evaluation of PERCY is only based on simulations. Moreover it considers single hop network. In reality, however, the actual performance of such techniques can only be evaluated using a real testbed, which can help uncover any technical issue that might surface as a result of realizing the proposed technique in real networks.

[1] This work is supported by Qatar National Research Fund (QNRF) No. 08-374-2-144.

$$\begin{pmatrix} L_1 \otimes L_2 \otimes L_3 \otimes L_4 \\ L_1 \otimes L_2 \otimes L_3 \\ L_1 \otimes L_2 \\ L_1 \end{pmatrix}$$

Equ. 1 Canonical Triangular Scheme

In this paper, we make the following contributions. First, we realize the PERCY algorithm [10] in the Qatar University testbed (QUMESH) [11] for single hops. Then, we extend the NC-based mechanism for multi-hop video delivery, providing insights as to how to maximize the perceived video quality based on individual link qualities. After that, we provide performance evaluation and a comparative study for both a single hop and multi-hop networks by considering different encoding and decoding strategies. Finally, we provide a trade-off study that addresses the compromise between complexity and received decoded layer gain for multi-hop NC on a mesh network testbed.

## II. RELATED WORKS

Video delivery over WMNs received intensive attention. The major works in the area of video delivery over multi-hop mesh networks can be categorized into two groups: non-NC-based and NC-based video delivery schemes. In the first group, prior works introduce many solutions for video transmission over WMNs. Within this group, review is focused on the path selection solution. One characteristic of multi-hop WMNs is that there are multiple routes between the source and destination nodes. Multi-path improves the efficiency of the network. When a link is broken because of channel quality or mobility there is an alternative path to transmit video to the destination, thus the E2E delay is decreased. Moreover, the throughput is increased as the data rate is allocated from different paths. When source nodes send MDC coding video to the destination, any packets are sent along any of the chosen path [17], while source nodes send base layer of MRC coding video with better selection of the path to enhance the network throughput [18]. Based on path quality [16], the source generates M sub-streams for the video, and then packets of sub-streams are transmitted among N paths. Finally, the destination reconstructs the video from received sub-streams. The multiple path transport (MPT) deals with two video code techniques, MRC and MDC [16]. To prevent base layer packet loss in MRC, the MPT retransmits lost base layers through enhancement layer paths, as without base layer packets in MRC, videos cannot be reconstructed. In contrast, in MPT with MDC, no retransmission is required, because videos can be reconstructed from any sub-stream with different quality and has no delay constraints. There are many challenges regarding path selection methods, however, finding the best disjoint paths and maintaining multiple paths to a destination require high complexity and cause overhead traffic over the network. One characteristic of WMNs that causes video quality degradation is wireless interference, thus to avoid interference among links, the paths must not be adjacent [15]. The similarity between MPT [16] and the framework is that video is generated and transmitted by a single source. In contrast to the framework in this paper, there is a static path, not multiple paths from source to destination. Moreover MPT is based only on MRC video techniques. To enhance the base layer transmission and provide high priority to base layer within framework in this paper, the base layer is encoded with every enhancement layers and retransmitted coded layers based on a canonical triangular scheme, which is similar to Koutsonikolas *et al.* [10].

NC is new paradigm, can improve video quality over non-NC category. In this category, the following works are reviewed. Halloush *et al.* proposed Multi Generation Mixing Network Coding (MGMNC) [6]. In MGM, packets are grouped into generations, and then MGM encodes packets of the same generation. MGMNC improves the reliability of delivering video and enhances decodable rates by using MGM combined with NC [6]. MGMNC supports layered video and the lower layers have a higher chance to deliver than the higher layers because the lower layers are combined with higher layers, which is similar to the framework within this study. Gheorghiu *et al.* [7] propose random linear network coding (RLNC) for layering video streams. The source and relay nodes know the IP address of the destination (similar to the approach in this study), thus the nodes do not know the entire network's topology or the encoding function, and complexity is decreased. The destination in RLNC [7] stores encoded packets in its buffer, waits for maximum delay in the video stream, then decodes by using Gaussian elimination. Subsequently, stores it in the playback buffer, finally it plays the video and sends minimal feedback to the source. RLNC [7] supports different video quality requirements for heterogeneous users. To decrease the complexity of the framework in this study, a triangular coding scheme is used instead of Gaussian elimination as in RLNC [7]. In the approach within this paper, NC is applied to layer packets from the same video at the single user, to increase the average useful decoded layer (AUDL) and to protect packet loss. Koutsonikolas et al. [10] propose the use of an NC mechanism called PERCY. To generate inter-layer NC for different video layers, a random linear coding is used to mix the $N$ packets from layer $i$, and the $N$ packets from layer $j$. To transmit layered video, the sender first estimates the one-hop PDR, and then selects the best NC mixing strategy based on the PDR of the link. Thereafter the sender generates coded packets based on the optimal strategy. However, PERCY utilises intra-session NC, not inter-session NC. The limitation of the approach in Koutsonikolas et al. [10] is that they use simulations to show a positive effect of using NC in a single hop WiFi scenario [8]. The simulation does not realise the full potential of the NC. The approach in this work is applied to a general multi-hop WMN, while PERCY is specific to single hop. Furthermore, an experiment approach based on real WMN test-bed is used to implement PERCY [10] to exploit the broadcast nature of the wireless medium, and network codes are selected based on PERCY to combine layers from the same video into a single code for transmission. Thus, AUDL is increased and the video quality is improved. Moreover, the performance of video is studied with a focus on video quality and delay. Although, PERCY is an optimal NC algorithm for single hop WMN, as the result of the framework shows, it is not optimal for multi-hop. Thus, decoding and re-encoding packets are proposed at the intermediate nodes to improve video quality.

The rest of the paper is organized as follows. In Section 4, experimental system model is described in details. The implementation of the NC-based layered video for single and multi-hop in QUMESH, and the obtained experimental results are described in Section 5. Finally the paper concludes and outlines the future works.

### III. SYSTEM MODEL

To evaluate video delivery over a WMN for single and multi-hop, a sender/receiver application is developed by implementing an Inter-Layer Network Coding algorithm. First the NC framework for a single hop network is described. Then, the NC concept for multi-hop networks is described in detail. After that, a simple NC threshold-based scheme is proposed for single and multi-hop networks. Finally, a performance comparison that highlights how the HBH NC scheme can overcome the limitations of realizing PERCY in multi-hop networks is conducted.

*A. NC mechanism design in a single hop network:*

By using the MRC technique, the sender divides video into GOPs, then the data is divided into $L$ layers, and each layer divided into a fixed number of packets $P$. The video is stored as a $2D$ array $L*P$. The total number of packets in video content is N [1][10]. To send layered video, a UDP connection is established between the sender and the receiver. Then the source node builds SPT once, at the beginning of the connection. Periodically, the source node computes the PDR value observed in the link connecting it to the destination node, based on a pre-configured number of probing packets.

The PDR value is used to select the optimal strategy to be considered in layered video transmission. The canonical coding scheme is used to send the layered video. At every updated period, the sender selects the best strategy according to the obtained PDR value. The transmission strategy is considered to transmit encoded video layers during the next updated period. Figure 2 shows the NC algorithm in the sender side.

*Algorithm NC_Sender (L,N)*
*Sender establishes UDP connection with Receiver*
*Build SPT*
*Send Probe PKT*
*LOOP (every updated period)*
  *Calculate PDR*
  *Select Best Strategy $(X_1, X_2, X_3, X_4)$ from SPT based on PDR according to the optimal scheme in [10]*
  *Encode and Send video based on best strategy*
*End LOOP*

Fig. 2 NC Sender Algorithm

*Algorithm NC_Receiver (L,N)*
*If connection is established successfully*
*LOOP*
  *IF PKT RCV then*
    *Num. of PKT RCV ++*
  *End IF*
  *Send Total Num. of PKT RCV successfully*
  *LOOP (For each PKT RCV)*
    *if $\sum_{j=i}^{i-k} L_j \geq (k+1), \forall k \in [0, i-1]$*
      *Decoded Layer = i*
  *End IF*
  *End LOOP*
 *Calculate Avg. Useful Layers Decoded*
*End LOOP*
*End IF*

Fig. 3 NC Receiver Algorithm

The receiver collects the coded packets that are successfully received. Based on the canonical triangular coding scheme, the receiver can decode the first $i$ layers $if \sum_{j=i}^{i-k} L_j \geq (k+1), \forall k \in [0, i-1]$ Equ. 2. For example, the receiver can decode the first 3 layers, $if L_3 \geq 1 \:\&\&\: L_3 + L_2 \geq 2 \:\&\&\: L_3 + L_2 + L_1 \geq 3$. It decodes the different layers based on the number of received packets. Then, it computes the number of useful layers that are received. Figure 3 summarizes the NC algorithm at the receiver side.

*B. NC mechanism design for a multi-hop network:*

In multi-hop network, we consider many intermediate nodes between source and destination nodes. We describe two different algorithms for NC-based layered video delivery. The first scheme considers only E2E encoding and decoding layered video, as the case of the NC mechanism described for single hop. In this scheme, the selection of the best transmission strategy is based on the E2E PDR metric. The second scheme considers HBH inter-layer video encoding and decoding strategies. While in E2E scheme the intermediate node only forward coded packets to the next hop, in HBH scheme, the intermediate node computes periodically the PDR of the link connecting it to the next hop. When the intermediate node receives the coded video packets, it decodes the different layers based on the number of received packets.

The selected transmission strategy is used to send inter-layer video packets and it is updated once the PDR of the link changes. Figure 4 shows the NC algorithm at the intermediate side.

---

*Algorithm NC_Intermediate (L,N)*
*IF connection is established successfully*
*Build SPT*
*Send Probing packet (PKT) to the following hop node*
*LOOP*
  *IF PKT RCV then*
    *Num. of PKT RCV ++*
  *End IF*
  *Send Total Num. of PKT RCV successfully to the previous hop node*
  *Calculate PDR*
  *Select Best Strategy from SPT based on PDR*
  *Forward video based on Best Strategy to the following hop node*
*End LOOP*
*End IF*

---

Fig. 4 NC Intermediate Algorithm

*C. NC threshold-based mechanism in a Single and a Multi-hop:*

Here we propose heuristic NC strategy, which uses simple threshold-based scheme to code and transmit layered video. The complexity of heuristic NC is $O(1)$, on the other hand, the complexity of PERCY is $O(L^2)$ [1]. Thus the main feature of heuristic is to reduce the complexity of algorithm. We define three threshold sets: in first set we define only one threshold, in second set we define two thresholds, and in third set we define three thresholds.

The metrics used to evaluate video performance in a wireless mesh network in single hop and multi-hop are:
- Number of Packets Received (NPR): The number of packets received successfully from the sender to the receiver.
- PDR: The total number of packets received successfully divided by the total number of packets sent.
- AUDL: The number of layers that can be decoded divided by the number of packets per layer. Our framework is emphasis on the application layer metrics such as AUDL that is improved the user reception video quality.

In the next section, multi-layers video delivery quality in single and multi-hop over the QUMESH test bed is examined. The benefits of the NC proposal and the different issues of real experimental evaluation are explored.

## IV. PERFORMANCE EVALUATION

The approach is experiment-driven, video delivery performance in the QUMESH testbed is performed for different scenarios. The testbed consists of 30 mesh nodes that are deployed indoor (although some mesh network links span outdoor structures) at Qatar University [11]. In this section we present the results of the experiments that have been performed to deliver inter layer-video over QUMESH. We describe the results of video quality in terms of AUDL and NPR metrics. We performed experiments over QUMESH testbed for three scenarios. In the first scenario, we performed our experiments in single hop and compared the obtained results with the results presented in Koutsonikolas *et al.* [10]. In the second scenario, we performed the experiments in a multi-hop network using three transmission schemes: E2E, HBH, and combination of E2E and HBH. First, we performed the multi-hop experiments in the E2E transmission scheme, then when all intermediate nodes are NC nodes. After that, we repeated the same experiments, when only some intermediate nodes are selected to be NC nodes. Finally, we compared the obtained results for the multi-hop scenario with the different proposed transmission scheme. In the third scenario, we performed our experiments by using a threshold-based scheme over single networks and multi-hop networks, the obtained results are compared with the results obtained from the first and second experiments. In our experiment, we considered the following parameters: *L* is equal to 4, *P* is equal to 8, and *N* is equal to 32. The video is stored as 2*D* array $L * P$.

*A. Evaluation for NC-based on a Single hop (NC1):*

The results of performing experiments over QUMESH when considering single hop are shown in Figure 5. The variation of AUDL as a function of PDR values observed during the experiments are presented, it was observed that the AUDL was better with the NC mechanism, compared to No-NC mechanism in different PDRs, especially when the PDR value was greater than 0.5. In contrast, when the PDR was between 0.25 and 0.5, the AUDL was lower with a NC mechanism compared to a No-NC. Although PERCY selects an optimal strategy for transmission, individual PDR values for links are computed periodically. Accordingly the pre-calculated PDR are used during the next transmission. However, the link quality could be changed during transmission over short time that effect in the obtained AUDL and video quality. This happened due to a lack of synchronization between the protocol and application layers. It was also found that when the loss rate was low, more than 50% of the layers were received successfully. The performance of the No-NC scheme decreased, when the PDR was low. By using a NC scheme, more video layers can be decoded because of the good link quality, which conforms to the results obtained by Koutsonikolas *et al.* [10], when the PDR value was greater than 0.5.

*B. Evaluation for NC-based on a Multi-hop (NC2, NC3):*

The benefit of enhancing the performance of video delivery in WMN in multi-hop is vital. Experiments are performed over QUMESH, considering a multi-hop scenario with the number of hops ranging from 1 to 3. By increasing the number of hops, the AUDL obtained with a No-NC mechanism is lower than the AUDL obtained with an NC mechanism, as shown in Figure 5. Moreover, Figure 5 illustrates the single hop case. It is evident that the AUDL with NC mechanism is better than the AUDL with NC mechanism in multi-hop. The gain of using NC diminishes by increasing the number of hops. This happens because there is significant difference between packets sent and packets received. As the number of hops is increased, NPR and PDR are low. We observed that the

AUDL is decreased to less than 20% - 30% compared to the single hop scenario. The E2E PDR metric is used to select the best strategy for the transmission video in single and multi-hop. In multi-hop, the E2E PDR does not capture the link

Figure 6 and Figure 7 show that the gain of NC2-HBH and NC3-HBH is better compared to NC2-E2E and NC3-E2E, respectively. Moreover, when the intermediate nodes use the NC mechanism, the AUDL outperforms that when it only forwards video. The packet loss is increased when intermediate nodes are increased between the sender and the receiver. Because of the AUDL as a function of PDR, calculating PDR per link increases the AUDL at the receiver. When intermediate nodes are not NC nodes and PDR is low, more packets are lost and the receiver cannot decode loss packets. In Figure 6, we also observe that NC2-HBH outperforms NC2-E2E by up to 10%, especially when the PDR is higher than 0.5. In Figure 7, we note that NC3-HBH outperforms NC3-E2E by up to 25% when the PDR is greater than 0.3. Moreover we observe that when the NC mechanism is applied in 1 node, the AUDL is increased by up to 20%. The improvement of adding 2 NC nodes instead of 1 NC node is not more than 5%, which might be considered insignificant, especially when we consider the delay involved in decoding, and re-encoding packets again using NC. The AUDL is an important metric for video quality and the delay is another important metric for video delivery.

Table 1 summarizes the average delay in single and multi-hop for E2E and HBH transmissions. The delay increases linearly with the number of NC-based intermediate nodes between the sender and the receiver. When only one intermediate node applies the NC mechanism, the delay is increased by 166% in 2 hops and 222% in 3 hops. The delay is increased by 545% when 2 NC intermediate nodes are used. In the NC node, the execution time of constructing SPT on QUMESH is on average 1 minute. At each NC node, the delay is increased because the processing time of SPT, the communication time, and the decoding and re-encoding times are increased. In NC3-HBH case, when using 1 NC node, the complexity and delay are less than using 2 NC nodes.

By these findings, a good trade-off would be to use NC at intermediate nodes that are part of low PDR links, while high PDR links may not necessarily use NC to mix packets, in order to address the trade-off between the AUDL performance gain and delay overhead as part of decoding and re-encoding at intermediate nodes.

TABLE I
AVERAGE DELAY (SECOND) IN SINGLE HOP & MULTI-HOP IN E2E & HBH

| #NC #Hop | E2E | 1 NC | 2 NC |
|---|---|---|---|
| 1 | 0.184 | - | - |
| 2 | 0.207 | 166.699 | - |
| 3 | 1.420 | 224.256 | 546,134 |

*C. Evaluation for heuristic NC in a Single and a Multi-hop:*

Here a simple threshold-based scheme is used to adjust the video transmission in both single and multi-hop networks. The PERCY algorithm was based on SPT. Although the processing quality, especially with different link qualities amongst different hops. Therefore, HBH PDR metric is the best to capture the link quality on each hop.

time for building SPT is large, it builds SPT at the beginning of communication before starting the transmission. To reduce the complexity of the PERCY and the delay of searching and selecting the optimal strategy to transmit coded video, static heuristic NC strategy based on the analysis of PERCY is presented. The results of the NC based on PERCY, were compared with the results obtained from the heuristic-NC in single hop and multi-hop networks. Experiments are performed, by using three threshold sets of PDR. In the first threshold set, the PDR was 0.5, if it is above 0.5 the strategy is (24,20,20,0), while when it is lower than threshold, the strategy is (64,0,0,0). In the second threshold set two thresholds were used, when the PDR is lower 0.3, between 0.3 and 0.8, and above 0.8, the strategies are (64,0,0,0), (48,16,0,0) and (24,20,20,0) respectively. In the third set, three thresholds were used, when the PDR is lower than 0.3, lower than 0.5, lower than 0.8, and above 0.8, the strategies are (64,0,0,0), (48,16,0,0), (24,20,20,0) and (40,8,8,8) respectively. Figures 8 and Figure 9 report the AUDL when using different heuristic strategies to send layered video over single and multi-hop networks. Figure 8 illustrates that AUDL with NC-PERCY outperforms AUDL with heuristic-NC over single hop networks, especially when PDR is greater than 0.55. This gain comes with a cost of delay and complexity. As shown in Figure 9, when nodes are sending video based on PERCY the resulted AUDL either is better or equal with heuristic strategy. This happened because of using a fixed strategy and this strategy was not based on the link quality but based on the static PDR threshold. The number of received packets did not increase proportionally by increasing the PDR value of the link. For example when the PDR is equal 1, the receiver can decode all layers, and there is no loss rate in link, but as observed in this study, only 3.3 layers are decoded in the case of third threshold set over 2 hops. By performing more experiments, changing thresholds and using different strategies, the AUDL can be increased but it is not easy to find a fixed optimal strategy. Although this difficulty, it comes with low complexity and delay.

V. CONCLUSION

In this framework, we realized PERCY on the QUMESH testbed. We proposed a new NC-based mechanism for delivering video over multi-hop networks. We have presented a detailed experimental study for video delivery over QUMESH for single and multi-hop wireless network. Our results show that NC-based schemes enhanced the video delivery in single and multi-hop scenarios, which the gain of using NC decreases by increasing the number of hops. By increasing the number of hops, the average useful decoded layers is decreased, but still more than 50% of layers are received successfully when the PDR is high. Introducing intermediate NC nodes can potentially increase the AUDL gain for multi-hop, which comes naturally at the cost of processing delay. However, we show that such intermediate NC-based schemes are not essential in all intermediate nodes. This calls for an opportunistic scheme that intelligently decides on incorporating an NC-based scheme as part of intermediate nodes to enhance performance, while minimizing the processing delay, and using wireless link states, which is a topic for future work.

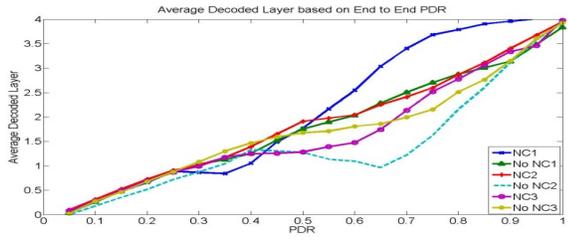

Fig. 5 Average decoded layers in single & multi-hop based on E2E NC

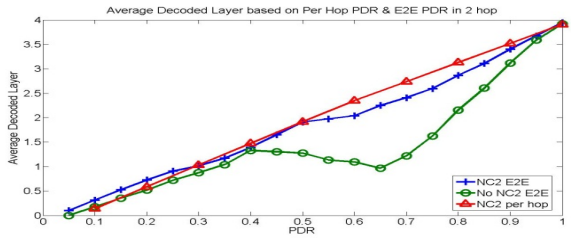

Fig. 6 Average decoded layers in 2 hops based on E2E & HBH NC

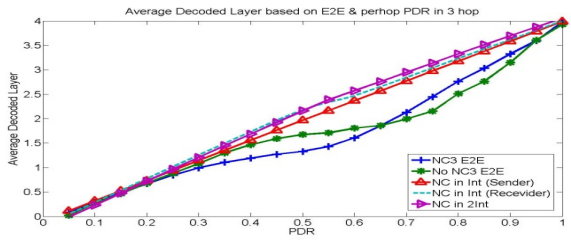

Fig. 7 Average decoded layers 3 hops based on E2E & HBH NC

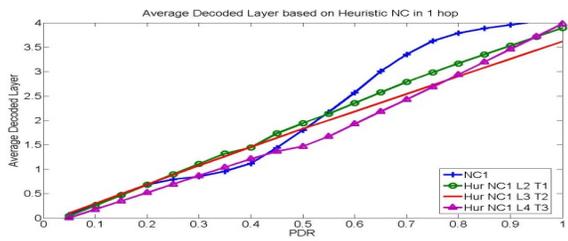

Fig. 8 Average decoded layer in single hop based on heuristic NC

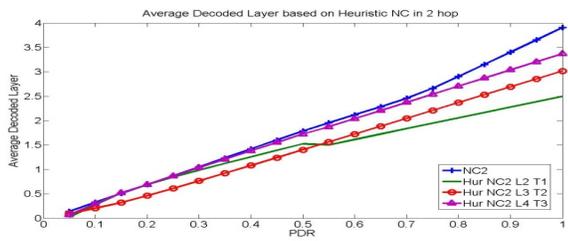

Fig. 9 Average decoded layers in 2 hops based on heuristic NC